\documentclass{aa}  
\def\gr{$\gamma$-ray}

\usepackage{graphicx}
\usepackage{txfonts}
\begin{document}

   \title{Microlensing constraint on the size of the gamma-ray emission region in blazar B0218+357}
   
   \author{Ie.Vovk\inst{1} \and A. Neronov\inst{2}}
   \institute{
     Max Planck Institut f\"ur Physik, F\"ohringer ring 6, 80805, Munich, Germany \\
     \and
     ISDC, Astronomy Department, University of Geneva, Ch. d'Ecogia 16, 1290, Versoix, Switzerland
   }
   
%


 
  \abstract
   {Observations of the effect of microlensing in gravitationally lensed quasars could potentially be used to study the structure of the source on distance scales down to the size of the supermassive black hole powering the quasar activity.}
   {We search for the microlensing effect in the \gr\ band using the signal from a gravitationally lensed blazar B0218+357.}
   {We develop a method of deconvolution of contributions of two images of the source  into the \gr\ band flaring lightcurve.  We use this method to study the evolution of the magnification factor ratio between the two images throughout the flaring episodes. We interpret the time variability of the ratio as a signature of the microlensing effect and derive constraints on the physical parameters of the \gr\ source by comparing the observed variability properties of the magnification factor ratio with those derived from  numerical simulations of the microlensing caustics networks.}
   {We find that the magnification factor ratio has experienced a change characteristic for a microlensing caustic crossing event during a 100~d flaring period  in 2012. It has further changed between 2012 and a recent flaring episode in 2014. We use the measurement of the maximal magnification and duration of the caustic crossing event to derive an estimate of the projected size of the $\gamma$-ray emission region in B0218+357, $R_\gamma\sim 10^{14}$~cm.  This estimate is compatible with a complementary estimate found from the minimal variability time scale. The microlensing / minimal variability time scale measurements of the source size suggest that the \gr\ emission is produced at the base of the blazar jet, in the direct vicinity of the central supermassive black hole.}
   {}


   \maketitle
%

\section{Introduction}
Almost hundred years after the discovery of the first large scale jet in an Active Galactic Nucleus (AGN) \citep{curtis18}, the origin of the jets and the details of the mechanism of their production and high-energy activity remain uncertain. 

Radio observations are now starting to reach the angular resolution sufficient to resolve the details of the region of the jet formation in the direct vicinity of the supermassive black hole(s) in nearby AGN \citep{doeleman12}. The angular resolution of  \gr\ telescopes will not be comparably high in any foreseeable future. This precludes the possibility to get the necessary imaging data  which would provide answers to the most basic  questions, such as e.g. the location of the site  of the $\gamma$-ray emission region in blazars \citep[AGNs with jets aligned along the line of sight,][]{urry95}. 

The only chance to overcome the limitation stemming from the limited angular resolution of  \gr\ telescopes is provided by the effect of the gravitational microlensing observable in the blazars / AGN which are parts of the strong gravitational lens systems \citep{chang79}. In such systems, the flux from small-scale sub-structures in the lensed source is selectively magnified by the effect of microlensing by individual stars in the lensing galaxy. The microlensing magnification factor  scales inversely proportional to the size of the source so that it affects mostly the flux from the smallest structures inside the source. Study of time and energy / wavelength dependence of the microlensing effect enables the study of the details of the source structure on the distance scales $R\ll 10^{16}$~cm (or micro-arcseconds), comparable to the size of the supermassive black hole powering the AGN activity \citep{torres,kochanek2004}. The microlensing data in the optical and X-ray continum and line emission allow to probe the structure of the accretion flow onto the black hole \citep{Courbin_disk_profiles,Dai_disk_profiles,Blackburne_disk_profiles,chartas12}, locate  the line emission regions within the AGN central engine \citep{microlensing_blr,guerras13} and/or the probe the properties of the lensing galaxy \citep{mediavilla15}. 

Only two strongly lensed blazars are detected in the GeV-TeV energy band by Fermi Large Area Telescope (LAT) \citep{1830_Fermi,B0218+35_Fermi} and by the MAGIC telescope \citep{B0218+35_2014_MAGIC}. The microlensing effect is observed in one of the two sources,  PKS 1830-211 \citep{neronov15}, limiting the source size to be not larger than 10-100 Schwarzschild radii of the supermassive black hole powering the source. 

The other \gr\ loud  blazar which is a part of the strong gravitational lens is  B0218+357 \citep{B0218+35_Fermi}. The system includes the lensing galaxy  at the redshift $z \approx 0.68$~\citep{B0218+35_lens_redshift} and a more distant blazar at $z \approx 0.94$~\citep{B0218+35_redshift}.  The gravitational time delay between the two images of the source is determined in the radio, $\tau_{rad}=10.5\pm 0.4$~d  \citep{Biggs_tdelay}, $\tau_{rad} = 10.1\pm 1.6$~d~\citep{Cohen_tdelay, Eulaers_tdelay} and in the  $\gamma$-ray band -- $\tau_\gamma \approx 11.46 \pm 0.16$~days~\citep{B0218+35_Fermi}.  The magnification factor ratio between the two images changes from $\mu_{radio}\simeq 2$ at 1.65~GHz frequency up to $\mu_{radio}\simeq 4$ at  15~GHz, presumably due to the free-free absorption in a giant molecular cloud  (GMC) located in the lensing galaxy in front of the image B \citep{mittal07}. 

In the \gr\ band, the two images of the source, separated by only $0.3''$, are not resolved. The magnification factor ratio is measured based on the fitting of the \gr\ lightcurve of the source, which is composed of the contributions from the two sources. This gives the result $\mu_\gamma\approx 1$~\citep{B0218+35_Fermi}, which is different from the free-free absorption corrected magnification factor ratio in the radio band~\citep{mittal07}. 

The difference in the magnification factor ratios between the radio and \gr\ bands could be a signature of the microlensing \citep{torres,1830_Fermi,B0218+35_Fermi,neronov15}. This would be interesting for constraining the size of the \gr\ source. However, the complicated chromatic behaviour of the radio magnification factor makes a straightforward interpretation of the magnification factor difference impossible. 

In what follows we notice that the microlensing does affect the magnification factor ratio in the \gr\ band. We deduce this from the time variability of the \gr\ magnification factor ratio \citep{B0218+35_Fermi,B0218+35_Fermi_paper_new}. Detection of the microlensing of the \gr\ flux of the source allows us to derive a constraint on the size of the \gr\ source. Similarly to the case of PKS 1830-211 \citep{neronov15}, the source size is found to be compact, with an order-of-magnitude size estimate  $\sim 10^{14}$~cm. The compact \gr\ source size indicates that the \gr\ emission is produced at the base of the jet, close to the supermassive black hole. 

\section{Fermi LAT data analysis}

In this work we use the publicly available Fermi/LAT Pass~7 reprocessed photon data set\footnote{http://fermi.gsfc.nasa.gov/cgi-bin/ssc/LAT/LATDataQuery.cgi}. The data were analysed using the \textit{Fermi Science Tools} package\footnote{http://fermi.gsfc.nasa.gov/ssc/data/analysis/scitools/} v9r33p0, using the ``Source'' (P7REP\_SOURCE) event class. The photons were selected from the $20^\circ$ region around the position of B0218+357. The fluxes of all sources in the selected region were estimated from the likelihood fit, which included all the sources from the 2FGL~\citep{2FGL} catalogue within 28 degrees from B0218+357.

The likelihood fitting procedure was repeated for each of the time bins of the source light curve. We used linear time binning of 0.5~day and extracted source fluxes in the 0.1-510~GeV energy band. We estimated the significance of detection in each time bin using the Test Statistics value~\citep[TS, ][]{mattox96}, obtained from the fit. Whenever the TS values dropped below 10 (which roughly corresponds to $3\sigma$ significance) the source detection was considered to be non-significant. 

\subsection{Timing analysis}

\begin{figure}
  \includegraphics[width=\linewidth]{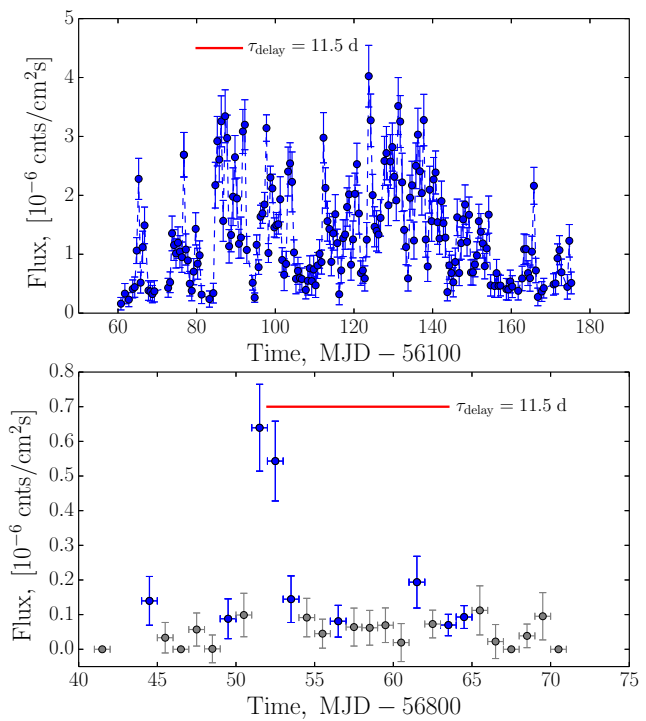}
  \caption{Light curves of the B0218+357 in 0.1-510~GeV band over its 2012 (top) and 2014 (bottom) flaring episodes. The bins size is 0.5~days for the 2012 and 1~day for 2014 light curve. Blue and grey data point mark the significant and non-significant detections correspondingly.}
  \label{fig::lc_comparison}
\end{figure}

Over the period of Fermi/LAT observations, B0218+357 has experienced two pronounced flaring periods~\citep{B0218+35_Fermi, B0218+35_Fermi_paper_new}. A zoom on these periods is shown in Fig.~\ref{fig::lc_comparison}.   The light curve of the source during the 2012 flare reveals strongly variable behaviour on different time scales, from the intraday to $\sim 100$~d \citep{B0218+35_Fermi}. The gravitational delay time scale is in between these minimal / maximal scales, so that a straightforward decomposition of the source flux onto the two separate contributions from the two images is not possible. 
Still, the gravitational time delay could be reliably estimated using the fast variability episodes occurring on the time scale much shorter than $\tau_\gamma$ \citep{B0218+35_Fermi}.  


\begin{figure}
 \includegraphics[width=\linewidth]{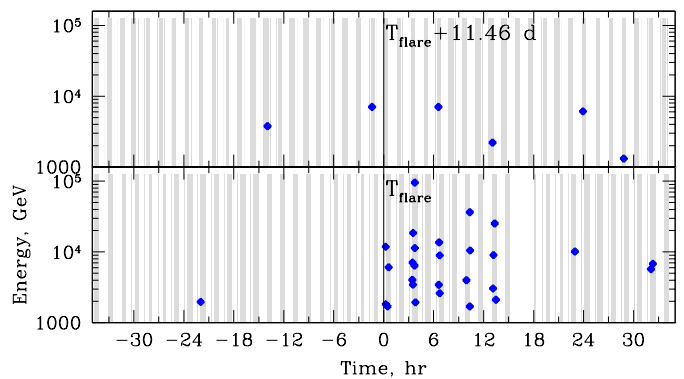}
  \caption{Arrival times and energies of photons above 1~GeV during the 2014 flare (bottom panel) and during the period after the gravitational time delay. Grey vertical bands mark the Good Time Intervals of Fermi / LAT. }
  \label{fig:lc_1GeV}
\end{figure}

 The gravitationally delayed flare expected 11.5~d after the 2014 episode is not detected. This is also clear from a detailed pattern of photon arrival times in the energy band above 1~GeV, shown in Fig. \ref{fig:lc_1GeV}. One could see that 24 photons are detected within a short 15~hr period of the main flare. The rise time of the flare is comparable to the time interval between subsequent exposures of the source by the LAT, $\simeq 3$~hr. Detection of the delayed flare with comparable flux and comparably sharp rising front would provide a precision measurement of the gravitational time delay with the minimal possible error bar of $3$~hr. However, no delayed flare is detected as it is clear from the upper panel of Fig. \ref{fig:lc_1GeV}, down to the flux level which is an order of magnitude lower than the flux of the main flare.
 
\subsection{Gamma ray magnification factor ratio}

Estimation of the magnification factor ratio between the two images of the source is difficult in the \gr\ band because the lensed images of the source are not resolved.  As a result, the  light curve of the \gr\ source has contributions from  both  images:
\begin{equation}
  F_{tot} = \mu_\gamma F(t)+F(t-\tau_\gamma)
  \label{eq::tot_flux}
\end{equation}
Here $\mu_\gamma$ denotes the magnification ratio of the leading image and $\tau_\gamma$ is the gravitational  time delay.

In real \gr\ data analysis, the continuous function $F(t)$ is replaced with a set of measurements of the source flux at discrete moments of time $F_i = F(t_i)$ and Eq.~\ref{eq::tot_flux} becomes:
\begin{equation}
  F^{tot}_i = \mu_\gamma F_i+F_{i-\Delta}
  \label{eq::tot_flux_discrete}
\end{equation}
where $\Delta$ denotes the shift $\tau_\gamma$ in terms of the light curve bins size. This equation can be re-written in the matrix form:
\begin{equation}
  \left|
  \begin{array}{c}
    F^{tot}_n      \\
    F^{tot}_{n-1}  \\
    \vdots         \\
    F^{tot}_{1}    \\
    F^{tot}_{base} \\
    \vdots         \\
    F^{tot}_{base} \\
    \cdots\\
  \end{array}
  \right|
  =  
  \left|
  \begin{array}{cccccccc}
    \mu_\gamma & 0          & \cdots     & 1          & 0      & \cdots     &        & 0  \\
    0          & \mu_\gamma & 0          & \cdots     & 1      & 0          & \cdots & 0  \\
    \vdots     &            &            &            &        &            &        &    \\
    0          & \cdots     & \mu_\gamma & 0          & \cdots & 1          & \cdots & 0  \\
    0          & \cdots     & 0          & \mu_\gamma & 0      & \cdots     & 1 & \cdots  \\
    \vdots     &            &            &            &        &            &        &    \\
    0          &            & \cdots     &            & \cdots & \mu_\gamma & 0      & 1 \\
    \cdots  &             & \cdots    &            &  & \cdots             &    \cdots    &    \\
  \end{array}
  \right|
  \left|
  \begin{array}{c}
    F_n      \\
    F_{n-1}  \\
    \vdots   \\
    F_{1}    \\
    F_{base} \\
    \vdots   \\
    F_{base} \\
    \cdots\\
  \end{array}
  \right|
  \label{eq::matrix_system}
\end{equation}

The source flux at the start of the observations ($t_{start}$) is composed of the total flux of the two images in the time interval preceding the $t_{start}$. In this way the system in Eq.~\ref{eq::matrix_system} can only be solved if the boundary condition for the moments of time $t<t_{start}$ is provided. In the  case of isolated flares one can assume that the source flux before the moment $t_{start}$ is equal to a constant quiescent flux $F_{base}$, as it is done in Eq. (\ref{eq::matrix_system}).

If $\tau_\gamma$ and $\mu_\gamma$ are known, an approximate solution of Eq.~\ref{eq::matrix_system} can be found from the least-square minimization under the condition $F_{i}>0$. The uncertainties of the flux estimates can be propagated to this system by simply dividing each value of $F^{tot}_{i}$ and $i$-th row of the matrix by the corresponding uncertainty $\sigma_i$. This transforms the least-square solution of the system in Eq.~\ref{eq::matrix_system} to the least-$\chi^2$ one.

The value of the time delay $\tau_\gamma$ can be found from the combined source lightcurve using e.g. the autocorrelation function analysis, which results in $\tau_\gamma \approx 11.5$~days~\citep{B0218+35_Fermi}. The magnification factor ratio $\mu_\gamma$ is more difficult to find because of the large number of uknowns $F_i$ in the system (\ref{eq::matrix_system}), equal to the number of equations. To assist the situation, an additional constraint on the solution $F_i$ can be imposed.

In the absence of accidental fine tuning of parameters, the true solution $F(t)$ of Eq.~\ref{eq::tot_flux} (or~Eq.~\ref{eq::tot_flux_discrete}) should not contain an autocorrelation at the time scale of the gravitational delay. In this way the best solution of the system in Eq.~\ref{eq::matrix_system} can be found by minimizing the residuals between the autocorrelation value of $F(t)$ around the position of expected time delay $\tau_\gamma$ and its local approximation, e.g. by a power law.

In a particular case of B0218+357, the major source flare in 2012 was preceded by the period of a relatively constant flux~\citep{B0218+35_Fermi}, which allows the estimation of the baseline flux $F_{base}$, needed to solve the system (\ref{eq::matrix_system}). 

Solving the system  (\ref{eq::matrix_system}) using the procedure described above and assuming $\mu_\gamma$ constant in time we find that no acceptable solution could be found. To get an acceptable solution, the assumption on the constancy of $\mu_\gamma$ has to be relaxed. In principle, the time scale of variability of $\mu_\gamma$ is not known a-priori. However, general considerations of microlensing suggest that even if the source size is as small as the size of the supermassive black hole, the variability time scale of the magnification factor ratio is the months-to-years range. 

\begin{figure}
  \includegraphics[width=\linewidth]{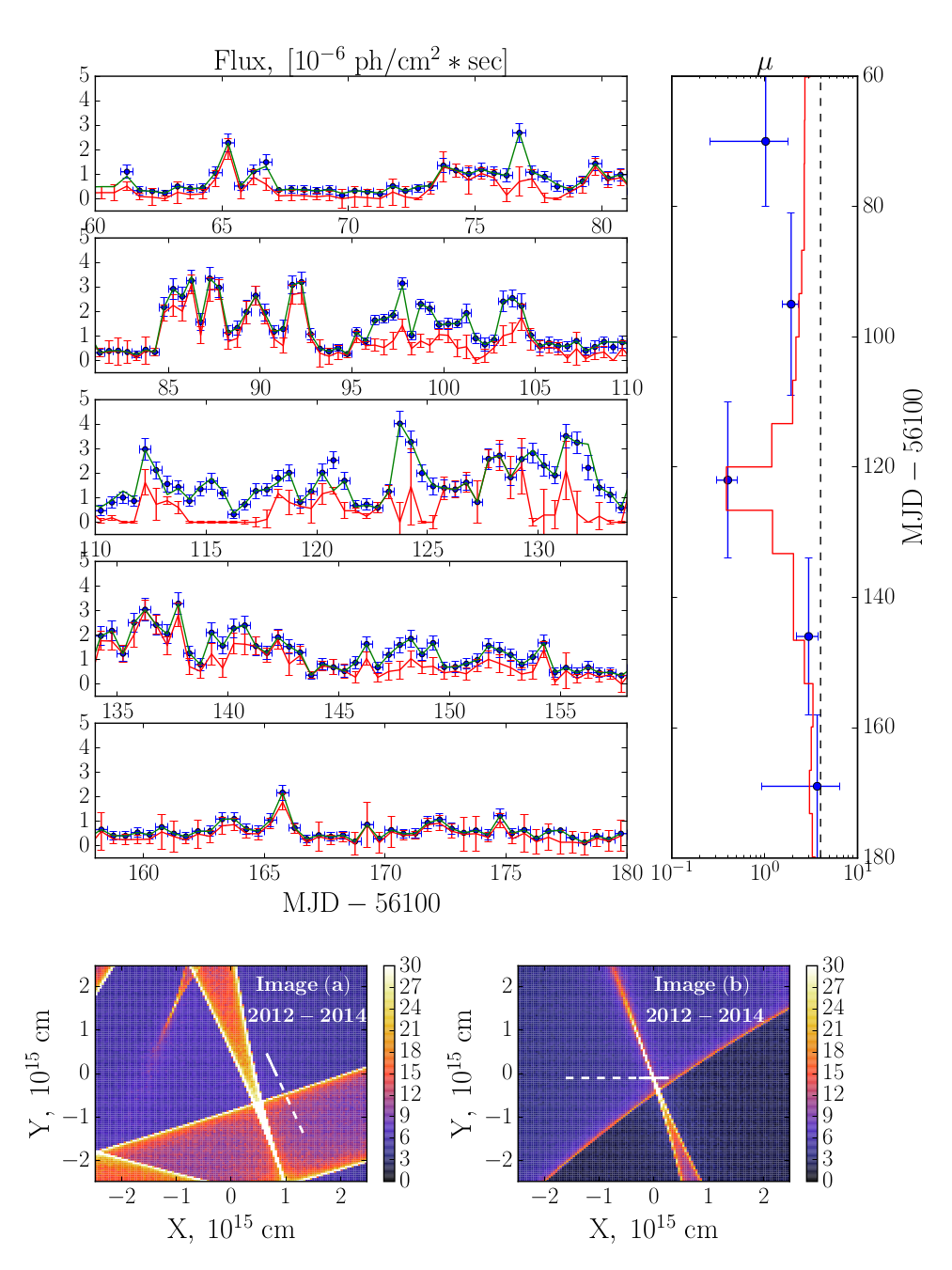}
  \caption{
    \textit{Top left:} Decomposition of the different flaring episodes of the B0218+357 major flare in 2012.
    \textit{Top right:} Derived magnification factor ratio for each of the flaring episodes. Black dashed line marks the approximate value of the magnification factor ratio in radio band $\mu_{rad} \approx 4$. The red line depicts the tentative association of the observed magnification factor behaviour with one the caustics crossing events in our simulations, computed for $R_{source} \approx 10^{14}$~cm. This event is shown in lower panels.
    \textit{Bottom:} Microlensing caustics magnification patterns, that can be associated with the detected variability of magnification factor over the 2012 and 2014 flares. Distances are given in the source plane. The patters are shown separately for the leading (left) and delayed images (right). Solid white lines mark the putative trajectories of the images during the 2012 flare, dashed lines depict the displacement of the images over the 2012-2014 period.
  }
  \label{fig::lc_decomposition}
\end{figure}

The 2012 major flare could be split on several shorter flaring episodes, identified by~\cite{B0218+35_Fermi}. To study the variability of $\mu_\gamma$, we modify our procedure for solving the system (\ref{eq::matrix_system}) by allowing $\mu_\gamma$ to change from sub-flare to sub-flare (in a step-wise manner). We search for a solution of the system (\ref{eq::matrix_system}) performing a scan over the magnification factor ratios for each flaring episode. This results in an acceptable solution $F(t)$ shown in Fig. \ref{fig::lc_decomposition}.

\begin{figure}
  \includegraphics[width=\linewidth]{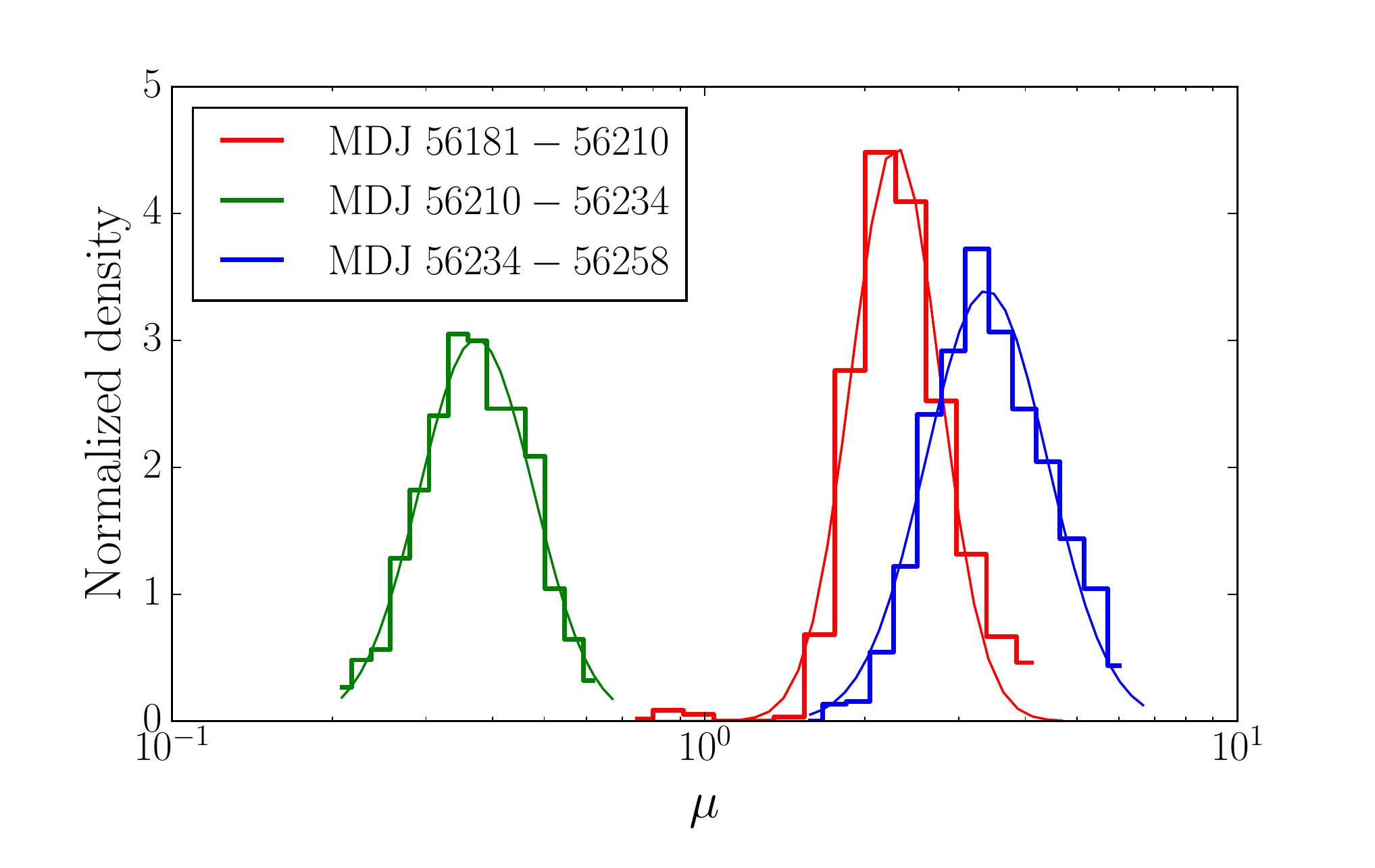}
  \caption{
   Distributions of the best-fit values of the magnification factor ratio $\mu$ for the three brightest flaring episodes in 2012 (3 central panels in Fig.~\ref{fig::lc_decomposition}), derived through the Monte Carlo simulations.
  }
  \label{fig::mu_fit}
\end{figure}

To estimate the uncertainties on the derived best-fit $\mu_\gamma$ for each flaring episode, as well as on the intrinsic lightcurve~$F(t)$, we have simulated a number artificial light curves $F_{tot}(t)$ in which the flux values in each time bin are randomly scattered within the error bars of the original flux measurement. These light curves were substituted to the system~\ref{eq::matrix_system} which was solved as described above. The distributions of the best-fit $\mu_\gamma$ for each flaring episode is shown in Fig. \ref{fig::mu_fit}. In order to quantify the scatter of the derived values of the magnification factor, we have fitted these distributions with a Gaussian profile. The width of the profiles is shown as the errorbars on the measurements of $\mu_\gamma$ in each episode in the right panel of Fig. \ref{fig::lc_decomposition}. 


We would like to note, that a similar behaviour of the magnification factor during the 2012 flare can be also seen in the analysis of~\cite{B0218+35_Fermi}, although the superposition of the leading and delayed flaring components did not allow there to make firm conclusions.

Contrary to~\cite{B0218+35_Fermi}, we do not find an evidence for the ``orphan'' sub-flares with no associated delayed emission. Detection of such sub-flares would imply a very rapid variation of the magnification factor, which would be interesting for the search of the microlensing. However, we find that the data are consistent with relatively slow modulation of the magnification factor ratio over the entire 2012 flare.

The magnification factor ratio, measured during the second flare of the source in 2014, shown in Fig.~\ref{fig::lc_comparison}, can not be directly established, as the source flux at the moment $T_{flare}+\tau_\gamma$ is consistent with the flux in-between the putative flares, i.e. the delayed emission was not detected by Fermi/LAT. The rising part of the  flare in the energy band above 1~GeV is very sharp, with the overall duration of several hours, see Fig. \ref{fig:lc_1GeV}. During the 15~hr flaring time, 24 photons were detected within the circle of the radius $0.5^\circ$ around the source. At the same time, only two or three photons were detected during the 15~hr time window of the delayed flare (compare the lower and upper panels of Fig. \ref{fig:lc_1GeV}). It is not clear to which of the two images belong the two / three photons detected during the delayed flare time window. The ratio of the photon counts of the ``prompt'' and ``delayed'' 2014 flares provides a lower bound on the magnification factor ratio $\mu_\gamma>5$ during this flare. 

\section{Discussion}

The analysis of the previous section shows that the magnification factor ratio $\mu_\gamma$ has varied during the 2012 flaring period and has further changed from 2012  to 2014.  An immediate consequence of the detection of time variability of $\mu_\gamma$ is the conclusion on the detection of the microlensing effect in the \gr\ band. An alternative possibility for the variations of the magnification factor ratio (as discussed by~\cite{Barnacka15} in the case of PKS 1830-211) would be different locations of the 2012 and 2014 flaring regions inside the source. This, however, would require the displacements of the order of the Einstein radius of the (macro)lens, $\sim 1$~kpc, even during one and the same flaring period. The scale of displacements is much larger than the typical estimates of the sizes of the \gr\ emission regions in blazars / quasars and radio galaxies and, moreover, in the case of the 2012 flaring episode, it is looks unlikely that two distant flaring regions would flare simultaneously within a 100~d period. 

On the other hand, the assumption of microlensing provides a natural explanation to the detected magnification factor ratio behaviour. Using the simulated micorlensing caustic patterns maps, described in more detail in section~\ref{sect::numerical_modelling}, we were able to tentatively associate the variation of $\mu_\gamma$ over the 2012-2014 flares with the event of caustic crossing of the delayed image in 2012 and a change in magnification of the leading image between 2012 and 2014 due to the movement over a larger scale pattern of lensing caustics. This tentative association is illustrated in the lower panels of Fig.~\ref{fig::lc_decomposition} and provides a reasonable description of the observed behaviour.

\subsection{Qualitative estimates of the microlensing constraint on the source size}

The microlensing magnifies the flux of the source if it is smaller or comparable in size to the lens Einstein radius:
\begin{equation}
  R_E(M) = \sqrt{\frac{4GM}{c^2} \frac{D_{LS}}{D_S D_L}} \approx 3 \times 10^{16} \left[\frac{M}{M_\odot}\right]^{1/2}\left[\frac{D}{1\mbox{ Gpc}}\right]^{-1/2}~\mathrm{cm}
  \label{eq::R_E}
\end{equation}
where $D_S,D_L,D_{SL}$ are the angular diameter distances to the source, the lens and between the source and the lens, $D$ is the overall distance scale and $M$ is the mass of the lens.

If the lensing centres in the lens galaxy are stars, only a source of the projected size less than $R_E$ corresponding to the lenses of the masses $\sim M_\odot$ is subject to the microlensing. In particular, the radio source which is known to be large (it is resolved in the radio band) $R_{radio}\sim 10$~pc \citep{mittal07} is not influenced by the microlensing by stars. 

The fact that the micorlensing effect is detected in the \gr\ band immediately implies a constraint on the size of the \gr\ source projected on the plane of the sky:
\begin{equation}
R_{\gamma,proj}\lesssim R_E(M_\odot)\simeq 3\times 10^{16}\mbox{ cm}
\end{equation}
under the ``default'' assumption $M\sim M_\odot$. 

The microlensing magnification factor is variable in time due to the relative motion of the source, the lensing centres and the observer.  Maximal magnification is achieved at the moments of crossing of the microlensing caustics in front of one of the two images of the source. At these moments, an order-of-magnitude estimate for the magnification factor is~\citep{chang84}:
\begin{equation}
  \mu^{micro} \simeq \sqrt{R_E/R_{\gamma,proj}}\simeq 10\left[\frac{R_{\gamma,proj}}{3\times 10^{14}\mbox{ cm}}\right]
  \label{eq::microlensing_magnification}
\end{equation}
Assuming that the high-frequency (15 GHz)  radio observations provide a fair estimate of the macro-lensing magnification factor ratio, $\mu_{radio}\simeq 3.7$, the overall magnification factor ratio is expected to vary in the range $\mu_{radio}/\mu^{micro}<\mu_\gamma<\mu_{radio}\mu^{micro}$, where we have made a simplifying assumption that both images of the source are affected by the microlensing in a similar way.

Measurement of the variability of the magnification factor ratio from $\mu_\gamma\lesssim 0.4\pm 0.1$ to $\mu_\gamma\gtrsim 5$ allows to make an estimate of $\mu_{micro}\sim 10$. This gives an oder-of-magnitude estimate of the source size
\begin{equation}
\label{eq:estimate}
R_{\gamma,proj}\sim 3\times 10^{14}\left[\frac{\mu^{micro}}{10}\right]^{-2}\mbox{ cm}
\end{equation}

An additional constraint on the projected size of the source stems from the measurement of the time scale of the variability of $\mu_\gamma$. The variability shown in the right panel of Fig.~\ref{fig::lc_decomposition} is characteristic for a caustic crossing episode during which $\mu^{micro}$ reaches a maximum / minimum. The typical timescale of the caustic crossing events is determined by the relative velocity $v$ of the source with respect to the caustics pattern and the observer~\citep{Kayser_ray_tracing}. Assuming that the velocities of the source and the lens are of the same order as the velocity of the Sun w.r.t. the Cosmic Microwave Background reference frame  $v\simeq 300$~km/s  \citep{kogut93} one could find that the strong magnification events related to the crossing of the source by the microlensing caustics last for 
\begin{equation}
  \Delta t_\gamma = R_{\gamma,proj} / v_{rel} \simeq 100\left[\frac{R_{\gamma,proj}}{3\times 10^{14}~\mathrm{cm}} \right] \left[ \frac{v_{rel}}{300~\mathrm{km/s}} \right]^{-1} ~\mathrm{d}
  \label{eq::microlensing_duration}
\end{equation}
which is consistent with the data of the 2012 flaring episode.

The caustic crossing also provides a natural explanation for the exceptional brightness of the 2012 flaring episode. In the absence of the microlensing, the flux from the dimmer image of the blazar is by a factor of three lower than the flux of the brighter component. A moderate flaring activity typically produces the flux at the level $F\sim 10^{-6}$~ph/(cm$^2$ s) in the brighter image of the source and just at the level $F\sim 3\times 10^{-7}$~ph/(cm$^2$ s) in the dimmer image. An order of magnitude change in the magnification factor of the dimmer image, caused by the caustic crossing resulted in a boost of the dimmer image flux up to  $F\sim 3\times 10^{-6}$~ph/(cm$^2$ s), so that the overall source flux has grown up to $F_{tot}\sim 4\times 10^{-7}$~ph/(cm$^2$ s) during the flaring period.

\subsection{Numerical modelling of the microlensing magnification factor statistics}
\label{sect::numerical_modelling}

Qualitative estimates presented above are uncertain because of the low statistics of the microlening events (caustic crossings) in the Fermi / LAT data. Only one such event is detected after six years of Fermi / LAT exposure. The maximal magnification factor and the time scale of the caustic crossing events fluctuate due to the complex shapes of the caustic patters. These fluctuations introduce an uncertainty into the estimates of the source size based on the microlensing data. 

\begin{figure}
  \includegraphics[width=\linewidth]{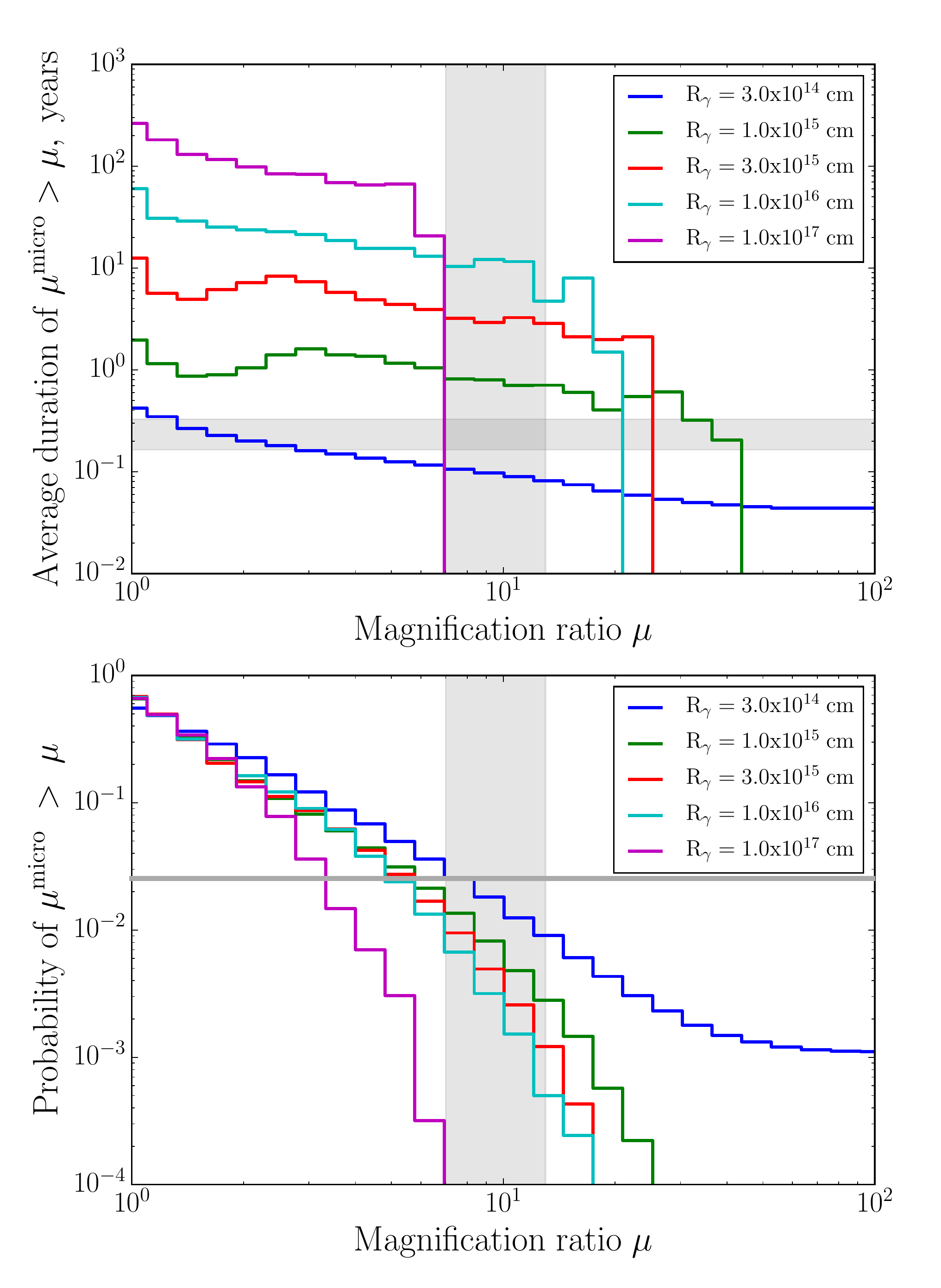}
  \caption{
    \textit{Top:} Distribution of the average durations of the microlensing magnification episodes obtained from the simulations for $\Sigma=0.5$ and $v=300$~km/s. Horizontal shaded region depicts the range $60-120$~days of the apparent variability time scales of the magnification factor ratio in B0218+357 and the vertical one depicts the microlensing magnification range, suggested by the observations.
    \textit{Bottom:} Cumulative probability of observing microlensing magnification $> \mu$ for different source sizes, estimated from the simulations. The vertical shaded region is the same as in the upper panel.
  }
  \label{fig::typical_durations}
\end{figure}

This uncertainty  is illustrated by Fig. \ref{fig::typical_durations} where the expected statistics of the magnification factor ratio with account of the microlensing effect  is calculated for different \gr\ source sizes. The calculation presented in this figure was performed via the simulation of the microlensing magnification maps using the inverse ray shooting method~\citep{Kayser_ray_tracing,Schneider_ray_tracing_86,Schneider_ray_tracing_87}. The simulated region of space encompasses 0.2x0.2~pc in the source plane and contains 300-1700 stars-lenses, depending on the assumed optical depth with respect to microlensing, which was scanned in range $\Sigma=[0.2,1.0]$. The lenses mass distribution was chosen to follow the~\citet{IMF_Chabrier} initial mass function with the cut-off at $10~M_{\odot}$. For each simulation run we have randomly selected 10 trajectories for each of the two lensed images of the source. The ``magnification light curves'', computed along these trajectories, represent the magnification of each of the images during the source movement with respect to the lens. In order to compare with the detected magnification factor ratio, we have computed the ratio of the trajectories for both images. Using these simulated magnification ratio curves, we then calculated the average durations of the episodes of high magnification $\mu^{micro}>\mu$.

Fig. \ref{fig::typical_durations} shows the dependence of the variability timescale of the microlensing on the source size, obtained from the simulations. One could see that a source of the size larger than $3\times 10^{15}$~cm is typically not expected to change the magnification factor ratio on the time scale shorter than two years, as observed in B0218+357 system. At the same time, a source of the size $3\times 10^{14}$~cm would exhibit variations of the magnification factor ratio already on the time scale of the 2012 flare. Thus, the detection of the microlensing implies an estimate of the source size $R_\gamma \lesssim 3 \times 10^{14} \left(v/300~\mathrm{km/s}\right)$~cm. 

The largest uncertainty of the estimate stems from the unknown projected velocity of the source. In principle, this velocity could be up to $v\sim c$, because the flaring source could be a relativistically moving blob inside the jet. Such fast motion of the flaring blob would explain the ``orphan'' high magnification sub-flares tentatively found in the analysis of \cite{B0218+35_Fermi}. At the same time, this would imply a significantly larger source size $R_\gamma \approx 3 \times 10^{17} (v/c) (\Delta t_\gamma / 100~\mbox{days})$~cm in order to account for the detected duration $\Delta t_\gamma \sim 100$~days of the high magnification event during the 2012 flaring episode. The possible apparent superluminal motion $v > c$ of the highly relativistic blobs would make this estimate even larger. Such source size is much larger than the size of the Einstein ring of the micro-lens in B0218+357. In this situation, the microlensing would not be able to significantly affect the source flux, as can be seen from the upper panel of Fig.~\ref{fig::typical_durations}. 
Still, a mildly relativistic source with $v\sim 0.1c$ of the size $\lesssim 10^{16}$~cm would already occasionally produce high-magnification events with $\mu^{micro}\sim 10$. 

The cases of large fast moving source and smaller slow moving source could be distinguished based on the expected statistics of the high-magnification events. During more than six years of Fermi / LAT observations, only one caustic crossing / high magnification event is detected. This fact provides a rough estimate of the probability to find the value $\mu^{micro}\gtrsim 10$ in the microlensing lightcurve, $p(\mu^{micro}\gtrsim 10)\simeq (60\mbox{ d}/ \left(6\cdot 365\mbox{ d}\right)\simeq 3$\%. Lower panel of Fig. \ref{fig::typical_durations} shows that this estimate favours the compact source of the size $\lesssim 3\times 10^{14}$~cm. If the source size is larger, the fraction of caustic crossing events with sufficiently high magnification decreases. Of course, with only one detected caustic crossing event, the estimate of $p(\mu^{micro}\gtrsim 10)\simeq 3\%$ is not very precise. Further monitoring of the source is needed to confirm the hypothesis of the small source size.  



\subsection{Comparison of the microlensing constraint with the constraint from the minimal variability time scale}

The microlensing constraint on the source size is consistent with the constraint stemming from the minimal variability time scale. The 2014 flare was particularly sharp in time, with the overall duration $\tau_{flare}\lesssim 15$~hr in the energy band above 1~GeV (see Fig. \ref{fig:lc_1GeV}). The rise time of the flare is comparable to the interval between the subsequent Fermi/LAT exposures of the source spaced by the $\tau_{rise}\simeq T_{LAT}=3.2$~hr period. This fast variability constrains the linear source size to be \citep{celotti,sibiryakov}
\begin{equation}
R_\gamma\lesssim c\tau_{rise}\Gamma^2 \simeq 3\times 10^{14}\Gamma^2\left[\frac{\tau_{rise}}{10^4\mbox{ s}}\right]\mbox{ cm}
\end{equation}
where $\Gamma$ is the bulk Lorentz factor of the jet. If the jet is aligned along the line of sight within a viewing angle $\theta\sim \Gamma^{-1}$, the size of the emission region projected on the plane of the sky is limited to 
\begin{equation}
R_{\gamma,proj}=R_\gamma\theta\lesssim c\tau_{rise}\Gamma\simeq 3\times 10^{15}\left[\frac{\Gamma}{10}\right]\left[\frac{\tau_{rise}}{10^4\mbox{ s}}\right]\mbox{ cm}
\end{equation}
Assuming the jet bulk Lorentz factor $\Gamma\sim 1-10$, one could find that the expected projected size of the \gr\ emission region derived from the variability analysis is comparable with the estimate of the size of the \gr\ emission region from the microlensing analysis. 

\subsection{Problem of escape of very-high-energy \gr s from the source}

Compactness of the \gr\ emission region in B0218+357 might pose a problem of escape of the very-high-energy \gr s, which have been recently detected by MAGIC telescope \citep{B0218+35_2014_MAGIC}. Gamma rays with energies above 100~GeV produce $e^+e^-$ pairs in interactions with UV photons of energies about $\epsilon\sim 10$~eV and, thus, might be significantly absorbed if the UV photon density is sufficiently big. 

Luminosity of the source in this energy range is about $L_{UV}\sim 10^{44}$~erg/s \citep{sed}. If this luminosity stems from the accretion flow, the UV photon distribution is isotropic in the black hole reference frame. If the UV source size is $R_{UV}$, the number density of the UV photons is about 
\begin{equation}
n_{ph}=\frac{L_{UV}}{4\pi R_{UV}^2 c \epsilon} \simeq 10^{11}\left[\frac{R_{UV}}{10^{16}\mbox{ cm}}\right]^{-2}
\left[\frac{L_{UV}}{10^{44}\mbox{ erg/s}}\right] \mbox{ cm}^{-3}
\end{equation}
The mean free path of the 100~GeV \gr s is limited to 
\begin{equation}
\lambda_{\gamma\gamma}=\frac{1}{\sigma_{\gamma\gamma}n_{ph}}\sim 10^{14}\left[\frac{R_{UV}}{10^{16}\mbox{ cm}}\right]^{2}\left[\frac{L_{UV}}{10^{44}\mbox{ erg/s}}\right]^{-1}\mbox{ cm}
\end{equation}
and the optical depth of the source with respect to the pair production is 
\begin{equation}
\tau_{\gamma\gamma}=\frac{R_{UV}}{\lambda_{\gamma\gamma}}\simeq 10^2\left[\frac{R_{UV}}{10^{16}\mbox{ cm}}\right]^{-1}\left[\frac{L_{UV}}{10^{44}\mbox{ erg/s}}\right]
\end{equation}
Thus, if the UV source is as compact as  the \gr\ source, $R_{UV}\lesssim 10^{15}$~cm, the highest energy ($100$~GeV) photons could not escape from the source. Detection of such photons by MAGIC telescope \citep{B0218+35_2014_MAGIC} imply that the UV source is large, 
\begin{equation}
R_{UV}\gtrsim 10^{18}\tau_{\gamma\gamma}^{-1}\left[\frac{L_{UV}}{10^{44}\mbox{erg/s}}\right]\mbox{ cm.}
\end{equation}

The origin of the source flux in the UV band is uncertain. The source spectrum does not exhibit a UV bump characteristic for the black hole accretion disk. If the emission is dominated by the jet, the bulk of the UV flux is most probably produced via inverse Compton scattering by relatively low energy electrons, so that the large size of the UV source does not appear surprising. Otherwise, if the accession flow onto the black hole in B0218+357 is radiatively inefficient, the UV emission is produced by the inverse Compton scattering by the non-relativistic electrons at large distances from the black hole. In this case the large size of the source is also naturally expected.

The constraint on the UV source size could be directly verified via the (non)-observation of the effect of the microlensing in the UV band. Indeed, a source of the size $R_{UV}\gtrsim R_E(M_\odot)$ should  be almost not affected by the microlensing. Simultaneous \gr\ and UV observations of the source could be used to measure the time variability of the magnification factor in the UV band (or the absence of it). Variations of the magnification factor ratio comparable to those observed in the \gr\ band should produce an order-of-magnitude variations of the UV flux. Tracing the variations of the magnification factor ratio via \gr\ observations and comparing them with the variations of the UV flux should provide a test of the presence / absence of the microlensing effect in the UV band. 

\section{Conclusions}

We have demonstrated that the \gr\ signal from the gravitationally lensed blazar B0218+357 is affected by the microlensing. Our conclusion is based on the detection of variability of the magnification factor ratio in the range $\mu_\gamma\simeq 0.4$ to $\mu_\gamma\gtrsim 5$ both on short ($\lesssim 100$~days) and long ($\sim 600$~days) time scales. 

Using the data on the time variability of the magnification factor ratio, we have derived a constraint on the projected size and location of the \gr\ emission region. The order-of-magnitude estimate of the size is $R_\gamma\sim 10^{14}-10^{15}$~cm. The microlensing estimate of the source size is consistent with an estimate stemming from the minimal variability time scale of the source.  The \gr\ emission region is not moving with relativistic speed so that it's most probable location is close to the blazar central engine, at the base of the jet. 

Constraints on the size and location of the \gr\ emission region in  B0218+357 are similar to those derived from the detection of microlensing in the other strongly lensed blazar PKS~1830-211 \citep{neronov15}. This demonstrates that the small size and location close to the supermassive black hole are generic features of \gr\ emission from blazars.

\begin{acknowledgements}
The work of I.V. is supported by the Swiss National Science Foundation grant P2GEP2\_151815.
\end{acknowledgements}

\bibliographystyle{aa}
\bibliography{bibliography}


\end{document}